\pdfoutput=1

\documentclass[
natbib,10pt]{TOOLS/sigplanconf}
\usepackage{amsmath}
\usepackage{graphicx}
\usepackage{subfig}
\usepackage{url}
\usepackage{verbatim} 
\usepackage{listings}
\lstnewenvironment{codex}
    {\lstset{}%
      \csname lst@SetFirstLabel\endcsname}
    {\csname lst@SaveFirstLabel\endcsname}
\lstnewenvironment{code}
    {\lstset{}%
      \csname lst@SetFirstLabel\endcsname}
    {\csname lst@SaveFirstLabel\endcsname}
\newtheorem{prop}{Proposition}
\newtheorem{df}{Definition}

\newcommand{\BI}[0]{\begin{itemize}}
\newcommand{\EI}[0]{\end{itemize}}

\newcommand{\BE}[0]{\begin{enumerate}}
\newcommand{\EE}[0]{\end{enumerate}}

\newcommand{\BX}[0]{\begin{codex}}
\newcommand{\EX}[0]{\end{codex}}
\newcommand{\BV}[0]{\begin{verbatim}}
\newcommand{\EV}[0]{\end{verbatim}}
    
\def \bscale1 {0.50}
\def \bscale {0.25}

\begin{document}

\title{
  
  
  Declarative Combinatorics: Boolean Functions, Circuit Synthesis and BDDs
  in Haskell }
\subtitle{-- unpublished draft --}
           
\authorinfo{Paul Tarau}
   {Department of Computer Science and Engineering\\
   University of North Texas}
   {\em tarau@cs.unt.edu}

\maketitle

\date{}

\begin{abstract}
We describe Haskell implementations of 
interesting combinatorial generation algorithms
with focus on boolean functions and
logic circuit representations.

First, a complete exact combinational logic circuit 
synthesizer is described
as a combination of {\em catamorphisms} and {\em anamorphisms}.

Using {\em pairing} and {\em unpairing} functions on natural
number representations of truth tables, we derive 
an encoding for Binary Decision Diagrams (BDDs) with the unique property 
that its boolean evaluation faithfully mimics
its structural conversion to a a natural number
through recursive application of a matching pairing function.

We then use this result to derive {\em ranking} and {\em unranking}
functions for BDDs and reduced BDDs.

Finally, a generalization of the encoding 
techniques to Multi-Terminal BDDs is
provided. 

The paper is organized as a self-contained literate Haskell program,
available at \url{http://logic.csci.unt.edu/tarau/research/2008/fBDD.zip}.

\end{abstract}

\section{Introduction}

This paper is an exploration with functional programming tools of {\em ranking}
and {\em unranking} problems on Binary Decision Diagrams. The practical
expressiveness of functional programming languages (in particular Haskell) 
are put at test in the process. The paper is part
of a larger effort to cover in a declarative programming 
paradigm, arguably more elegantly, some fundamental combinatorial generation 
algorithms along the lines of \cite{knuth06draft}.

The paper is organized as follows:

Sections \ref{bits} and \ref{bdds} overview efficient evaluation of boolean
formulae in Haskell using bitvectors represented as arbitrary length integers
and Binary Decision Diagrams (BDDs).
 
Section \ref{circ} describes an exact combinational circuit synthesizer.
 
Section \ref{pairings} discusses classic pairing and unpairing
operations and introduces new pairing/unpairing
functions acting directly on bitlists.

Section \ref{encbdd} introduces a novel BDD encoding (based on our unpairing
functions) and discusses the surprising equivalence between boolean evaluation of BDDs
and the inverse of our encoding, the main result of the paper.

Section \ref{rank} describes {\em ranking} and {\em unranking}
functions for BDDs and reduced BDDs.

Section \ref{multi} extends our results to Multi-Terminal BDDs.

Sections \ref{related} and \ref{concl} discuss related work, 
future work and conclusions.

The code in the paper, embedded in a literate programming LaTeX
file, is entirely self contained and has been tested under {\tt GHC 6.4.3}.

\section{Evaluation of Boolean 
Functions with Bitvector Operations}\label{bits}

Evaluation of a boolean function can be performed one 
bit at a time as in the function {\tt if\_then\_else}
\begin{code}
if_then_else 0 _ z = z
if_then_else 1 y _ = y
\end{code}
\noindent resulting in
\begin{codex}
> [([x,y,z],if_then_else x y z)|
      x<-[0,1],y<-[0,1],z<-[0,1]]
  [([0,0,0],0),
   ([0,0,1],1),
   ([0,1,0],0),
   ([0,1,1],1),
   ([1,0,0],0),
   ([1,0,1],0),
   ([1,1,0],1),
   ([1,1,1],1)]   
\end{codex}
Clearly, this does not take advantage of the ability of modern hardware to
perform such operations one word a time - with the instant benefit of a
speed-up proportional to the word size.
An alternate representation, adapted
from \cite{knuth06draft} uses integer encodings 
of $2^n$ bits for each boolean variable $x_0,\ldots,x_{n-1}$. 
Bitvector operations are used to evaluate all
value combinations at once.

\begin{prop}
Let $x_k$ be a variable for $0 \leq k<n$
where $n$ is the number of distinct variables in a 
boolean expression. Then column $k$ of the truth table
represents, as a bitstring, the natural number:

\begin{equation} \label{var}
x_k={(2^{2^n}-1)}/{(2^{2^{n-k-1}}+1)} 
\end{equation}
\end{prop}

\noindent For instance, if $n=2$, the formula computes 
$x_0=3=[0,0,1,1]$ and $x_1=5=[0,1,0,1]$.

The following functions, working with arbitrary length bitstrings are
used to evaluate the [0..n-1] variables $x_k$ with formula \ref{var} and map the
constant {\tt 1} to the bitstring of length $2^n$, {\tt 111..1}:
\begin{code}
-- the k-th, out of n bitvector boolean variables
var_n n k = var_mn (bigone n) n k

-- the k-th, out of n boolean variables w.r.t mask
var_mn mask n k = mask `div` (2^(2^(n-k-1))+1)

-- represents constant 1 as 11...1
bigone nvars = 2^2^nvars - 1
\end{code}

We have used in {\tt var\_n} an adaptation of the efficient 
bitstring-integer encoding described in the Boolean Evaluation 
section of \cite{knuth06draft}. Intuitively, it is based on the idea that one
can look at $n$ variables as bitstring representations of the $n$ columns
of the truth table.

Variables representing such bitstring-truth tables 
(seen as {\em projection functions}) 
can be combined with the usual bitwise integer operators, 
to obtain new bitstring truth tables, 
encoding all possible value combinations of their arguments.
Note that the constant $0$ is represented as $0$ while the constant $1$
is represented as $2^{2^n}-1$, corresponding to a column in
the truth table containing ones exclusively.

\section{Exact Combinational Circuit Synthesis} \label{circ}

A first application of 
these variable encodings is
combinational circuit synthesis, known to be intractable 
for anything beyond a few input variables.
Clearly, a speed-up by a factor proportional to the
machine's wordsize matters in this case.

\subsection{Encoding the Primary Inputs}
First, let us extend the encoding to cover constants {\tt 1} and {\tt 0},
that we will represent as ``variables'' {\tt n} and {\tt n+1} and
encode as vectors of {\tt n} zeros or {\tt n} ones (i.e. $2^{2^n}-1$,
passed as the precomputed parameter {\tt m} to avoid costly recomputation).
\begin{code}
encode_var m n k | k==n = m
encode_var m n k | k==n+1 = 0
encode_var m n k = var_mn m n k
\end{code}

Next we can precompute all the inputs knowing the number n of
primary inputs for the circuit we want to synthesize:
\begin{code}
init_inputs n = 
  0:m:(map (encode_var m n) [0..n-1]) where 
  m=bigone n
\end{code}
\begin{codex}
>init_inputs 3
  [0,15,3,5]
>init_inputs 3
  [0,255,15,51,85]
\end{codex}

Given that inputs have all distinct encodings, we can decode them
back - this function will be needed after the circuit is found.
\begin{code}
decode_var nvars v | v==(bigone nvars) = nvars
decode_var nvars 0 = nvars+1
decode_var nvars v = head 
  [k|k<-[0..nvars-1],(encode_var m nvars k)==v] 
    where m=bigone nvars
\end{code}
\begin{codex}
>map (decode_var 2) (init_inputs 2)
  [3,2,0,1]
>map (decode_var 3) (init_inputs 3)
  [4,3,0,1,2]
\end{codex}

We can now connect the inputs to their future occurrences
as leaves in the tree representing the circuit. This
means simply finding all the functions from
the set of inputs to the set of occurrences,
represented as a list (with possibly repeated)
values of the inputs.
\begin{code}
bindings 0 us = [[]]
bindings n us = 
  [zs|ys<-bindings (n-1) us,zs<-map (:ys) us]
\end{code}
\begin{codex}
>bindings 2 [0,3,5]
  [[0,0],[3,0],[5,0],[0,3],[3,3],
   [5,3],[0,5],[3,5],[5,5]]
\end{codex}
For fast lookup, we place the precomputed value
combinations in a list of arrays.
\begin{code}
generateVarMap occs vs  =
  map (listArray (0,occs-1)) (bindings occs vs)
\end{code}
\begin{codex}
>generateVarMap 2 [3,5]
  [array (0,1) [(0,3),(1,3)],
   array (0,1) [(0,5),(1,3)],
   array (0,1) [(0,3),(1,5)],
   array (0,1) [(0,5),(1,5)]]
\end{codex}

\subsection{The Folds and the Unfolds}

We are ready now to generate trees
with library operations marking
internal nodes of type {\tt F}
and primary inputs marking the leaves of type {\tt V}.
\begin{code}
data T a = V a | F a (T a) (T a) deriving (Show, Eq)
\end{code}
Generating all trees is a variant of an {\tt unfold} operation ({\em
anamorphism}).
\begin{code}
generateT lib n = unfoldT lib n 0
        
unfoldT _ 1 k = [V k]
unfoldT lib n k = [F op l r | 
  i<-[1..n-1], 
  l <- unfoldT lib i k, 
  r <- unfoldT lib (n-i) (k+i),
  op<-lib]
\end{code}
For later use, we will also define the dual {\tt fold} operation
({\em catamorphism}) parameterized by a function {\tt f} describing
action on the leaves and a function {\tt g} describing action on
the internal nodes. 
\begin{code}
foldT _ g (V i) = g i
foldT f g (F i l r) = 
  f i (foldT f g l) (foldT f g r)
\end{code}
This catamorphism will be used later in the synthesis process
for things like boolean evaluation. A simpler use would be to
compute the size of a formula as follows:
\begin{code}
fsize t = foldT f g t where
   g _ = 0
   f _ l r = 1+l+r
\end{code}
A first use of {\tt foldT} will be to decode the constants
and variables occurring in the result:
\begin{code}
decodeV nvars is i = V (decode_var nvars (is!i))

decodeF i x y = F i x y

decodeResult nvars (leafDAG,varMap,_) = 
  foldT decodeF (decodeV nvars varMap) leafDAG
\end{code}
The following example shows the action of the decoder:
\begin{codex}
>decodeV 2 (array (0,1) [(0,5),(1,3)]) 0
  V 1
>decodeV 2 (array (0,1) [(0,5),(1,3)]) 1
  V 0
>decodeResult 2 ((F 1 (V 0) (V 1)), 
            (array (0,1) [(0,5),(1,3)]), 4)
  F 1 (V 1) (V 0)
\end{codex}
The following function uses {\tt foldT} to generate
a human readable string representation of the result
(using the {\tt opname} function given in Appendix):
\begin{code}
showT nvars t = foldT f g t where
  g i = 
     if i<nvars 
       then "x"++(show i) 
       else show (nvars+1-i)
  f i l r =(opname i)++"("++l++","++r++")" 
\end{code}
\begin{codex}
> showT 2 (F 4 (V 0) (F 1 (V 1) (V 0)))
  "xor(x0,nor(x1,x0))"
\end{codex}

\subsection{Assembling the Circuit Synthesizer}
A Leaf-DAG generalizes an ordered tree by
fusing together equal leaves. Leaf equality in
our case means sharing a primary input 
variable or a constant.

In the next function 
we build candidate Leaf-DAGs by combining two
generators: the inputs-to-occurrences generator
{\tt generateVarMap} and the expression tree generator
{\tt generateT}. Then we compute their bitstring
value with a {\tt foldT} based boolean formula evaluator.
The function is parameterized by a library of logic gates {\tt lib},
the number of primary inputs {\tt nvars} and the maximum
number of leaves it can use {\tt maxleaves}:
\begin{code}
buildAndEvalLeafDAG lib nvars maxleaves = [
  (leafDAG,varMap,
     foldT (opcode mask) (varMap!) leafDAG) |
       k<-[1..maxleaves],
       varMap<-generateVarMap k vs,
       leafDAG <-generateT lib k
  ] where
      mask=bigone nvars
      vs=init_inputs nvars
\end{code}
We are now ready to test if the candidate matches
the specification given by the truth table of
{\tt n} variables {\tt ttn}.
\begin{code}
findFirstGood lib nvars maxleaves ttn = 
  head [r|r<-
    buildAndEvalLeafDAG lib nvars maxleaves,
    testspec ttn r
  ] where
  testspec spec (_,_,v) = spec==v
\end{code}
\begin{codex}
> findFirstGood [1] 2 8 1
  (F 1 (F 1 (V 0) (V 1)) (F 1 (V 2) (V 3)),
    array (0,3) [(0,5),(1,0),(2,3),(3,0)],1)
\end{codex}
The final steps of the circuit synthesizer
consist in converting to a human readable form
the successful
first candidate (guaranteed to be minimal
as they have been generated by increasing order of nodes).

\begin{code}
synthesize_from lib nvars maxleaves ttn = 
  decodeResult nvars candidate where
  candidate=findFirstGood lib nvars maxleaves ttn

synthesize_with lib nvars ttn = 
  synthesize_from lib nvars (bigone nvars) ttn

-- synthesizes an shows a circuit
syn lib nvars ttn = 
  (show ttn)++":"++
  (showT nvars (synthesize_with lib nvars ttn))

-- shows all circuits synthesized 
-- for functions with nvars inputs
synall lib nvars = 
  map (syn lib nvars) [0..(bigone nvars)]
\end{code}
The following example shows a minimal circuit
for the {\tt 2} variable boolean function with truth table
{\tt 6} ({\tt xor}) in
terms of the library with opcodes in {\tt [0]} i.e.
containing only the operator {\tt nand}. Note that
codes for functions represent their truth tables
i.e. {\tt 6} stands for  {\tt [0,1,1,0]}.
\begin{codex}
> syn [0] 2 6
  "6:nand(nand(x0,nand(x1,1)),nand(x1,nand(x0,1)))"
\end{codex}
The following examples show circuits synthetized for 3 argument
function {\tt if-the-else} in terms of a few different libraries.
As this function is the building block of boolean circuit
representations like Binary Decision Diagrams, having {\em perfect}
minimal circuits for it in terms of a given library has clearly
practical value. The reader might notice that it is quite unlikely
to come up intuitively with some of these synthesized circuits.
\begin{codex}
> syn symops 3 83
  "83:nor(nor(x2,x0),nor(x1,nor(x0,0)))"
>syn asymops 3 83
  "83:impl(impl(x2,x0),less(x1,impl(x0,0)))"
>syn mixops 3 83
  "83:nand(impl(x2,x0),nand(x1,x0))"
> syn [3,4] 3 83
  "83:xor(x1,less(xor(x2,x1),x0))"
\end{codex}
We refer to the Appendix for a few details,
related to the bitvector operations on various
boolean functions used in the libraries,
as well as a few tests.

\section{Binary Decision Diagrams} \label {bdds}

We have seen that Natural Numbers in $[0..2^{2^n}-1]$ can be used as
representations of truth tables defining $n$-variable boolean functions.
A binary decision diagram (BDD) \cite{bryant86graphbased} is an ordered binary 
tree obtained from a boolean function, by assigning its variables, one at a time, 
to {\tt 0}  (left branch) and {\tt 1} (right branch). 

The construction is known as Shannon expansion \cite{shannon_all}, and is
expressed as a decomposition of a function in two {\em cofactors}, $f[x
\leftarrow 0]$ and $f[x \leftarrow 1]$

\begin{equation}
f(x)= (\bar{x} \wedge f[x \leftarrow 0]) \vee (x \wedge f[x \leftarrow 1])
\end{equation}

\noindent where $f[x \leftarrow a]$ is computed 
by uniformly substituting $a$ for $x$ in $f$. Note that by using the more
familiar boolean {\tt if-the-else} function, the Shannon expansion can also
 be expressed as:

\begin{equation}
f(x) = if~x~then~f[x \leftarrow 0]~else~f[x \leftarrow 1]
\end{equation}

Alternatively, we observe that the Shannon expansion
can be directly derived from a $2^n$ size truth table, 
using bitstring operations on encodings of its $n$ variables.
Assuming that the first column of a truth table corresponds to 
variable $x$, $x=0$ and $x=1$ mask out, respectively, 
the upper and lower half of the truth table.

{\em Seen as an operation on bitvectors, the Shannon expansion 
(for a fixed number of variables) defines a bijection associating 
a pair of natural numbers (the cofactors's truth tables) 
to a natural number (the function's truth table), 
i.e. it works as a {\em pairing function}.}

\section{Pairing Functions} \label{pairings}

\begin{df}
A {\em pairing function} is a bijection $f : Nat \times Nat \rightarrow
Nat$. An {\em unpairing function} is a bijection $g : Nat \rightarrow
Nat  \times Nat$.
\end{df}

\subsection{Classic Pairing Functions}

Following Julia Robinson's notation \cite{robinson50}, 
given a pairing function $J$, its left and right inverses $K$ and $L$ 
are such that

\begin{equation}
J(K(z),L(z))=z
\end{equation}

\begin{equation}
K(J(x,y))=x
\end{equation}

\begin{equation} 
L(J(x,y))=y 
\end{equation}

We refer to  \cite{DBLP:journals/tcs/CegielskiR01} for a typical use 
in the foundations of mathematics and to \cite{DBLP:conf/ipps/Rosenberg02a} 
for an extensive study of various pairing functions and their computational properties. 

Starting from Cantor's pairing function
\begin{equation}
f(x,y)=(x+y)*(x+y+1)/2 + y
\end{equation}
and the Pepis-Kalmar-Robinson function
\begin{equation}
f(x,y)=2^x*(2*y+1)-1
\end{equation}
bijections from $Nat \times Nat$ to $Nat$ have been used for various proofs 
and constructions of mathematical objects 
\cite{pepis,kalmar1,robinson50,robinson55,robinsons68b,DBLP:journals/tcs/CegielskiR01}.

\subsection{Pairing/Unpairing 
operations acting directly on bitlists} \label{BitMerge}

We will introduce here a  pairing function, 
expressed as simple bitlist transformations.
This unusually simple pairing function (that we
have found out recently as being the same as the one in defined 
in Steven Pigeon's PhD thesis on Data Compression \cite{pigeon}, page 114),
provides compact representations for various constructs involving ordered pairs.

The function {\tt bitmerge\_pair} implements a bijection from $Nat \times
Nat$ to $Nat$ that works by splitting a number's big endian bitstring
representation into odd and even bits, while its inverse {\tt bitmerge\_unpair}
blends the odd and even bits back together. The helper functions 
{\tt nat2set} and {\tt set2nat}, 
given in the Appendix, convert from/to 
natural numbers to sets of nonzero bit positions.

\begin{code}
bitmerge_pair (i,j) = 
  set2nat ((evens i) ++ (odds j)) where
    evens x = map (2*) (nat2set x)
    odds y = map succ (evens y)
  
bitmerge_unpair n = (f xs,f ys) where 
  (xs,ys) = partition even (nat2set n)
  f = set2nat . (map (`div` 2))
\end{code}

The transformation of the bitlists
is shown in the following example with bitstrings aligned:
\begin{codex}
>bitmerge_unpair 2008
  (60,26)

-- 2008:[0, 0, 0, 1, 1, 0, 1, 1, 1, 1, 1]
--   60:[      0,    1,    1,    1,    1]
--   26:[   0,    1,    0,    1,    1   ]
\end{codex}

\begin{prop}
The following function equivalences hold:

\begin{equation}
bitmerge\_pair \circ bitmerge\_unpair \equiv id
\end{equation}

\begin{equation}
bitmerge\_unpair \circ bitmerge\_pair \equiv id
\end{equation}
\end{prop}

\section{Pairing Functions and Encodings of Binary Decision Diagrams}
\label{encbdd}

We will build a $BDD$ by applying {\tt bitmerge\_unpair}
recursively to a Natural Number {\tt tt}, 
seen as an $n$-variable $2^n$ bit truth table. 
This results in a complete binary tree of depth $n$.
As we will show later, this binary tree represents
a $BDD$ that returns {\tt tt} when evaluated applying
its boolean operations.

We represent a $BDD$ in Haskell as a binary tree {\tt BT} with constants {\tt 0}
and {\tt 1} as leaves, marked with the function symbol {\tt C}. Internal
nodes representing {\tt if-then-else} decision points, marked with {\tt D}, are
controlled by variables, ordered identically in each branch,
as first arguments of {\tt D}. 
The two other arguments are subtrees representing the {\tt THEN} 
and {\tt ELSE} branches. Note that, in practice, reduced, 
canonical DAG representations are used instead of
binary tree representations.

\begin{code}
data BT a = C a | D a (BT a) (BT a) 
             deriving (Eq, Show)
\end{code}

The constructor BDD wraps together the number of variables of
a binary decision diagram and the binary tree representation it.

\begin{code}
data BDD a = BDD a (BT a) deriving (Eq, Show)
\end{code}

The following functions apply {\tt bitmerge\_unpair} recursively, 
on a Natural Number {\tt tt}, seen as an $n$-variable $2^n$ bit truth table, 
to build a complete binary tree of depth $n$, 
that we will represent using the BDD data type. 

\begin{code}
-- n=number of variables, tt=a truth table
plain_bdd n tt = BDD n bt where 
  bt=if tt<max then shf bitmerge_unpair n tt
     else error 
       ("plain_bdd: last arg "++ (show tt)++
       " should be < " ++ (show max))
     where max = 2^2^n

-- recurses to depth n, splitting tt into pairs
shf f n tt | n<1 =  C tt 
shf f n tt = D k (shf f k tt1) (shf f k tt2) where
  k=pred n
  (tt1,tt2)=f tt
\end{code}
The following examples 
show the results returned by {\tt plain\_bdd} 
for all $2^{2^n}$ truth tables associated to $n$ variables
for $n=2$, with help from printing function {\tt print\_plain} given
in Appendix.

\begin{codex}
>print_plain 2
 BDD 2 (D 1 (D 0 (C 0) (C 0)) (D 0 (C 0) (C 0)))
 BDD 2 (D 1 (D 0 (C 1) (C 0)) (D 0 (C 0) (C 0)))
 BDD 2 (D 1 (D 0 (C 0) (C 0)) (D 0 (C 1) (C 0)))
 ...
 BDD 2 (D 1 (D 0 (C 0) (C 1)) (D 0 (C 1) (C 1)))
 BDD 2 (D 1 (D 0 (C 1) (C 1)) (D 0 (C 1) (C 1)))
\end{codex}

\subsection{Reducing the $BDDs$}
The function {\tt bdd\_reduce} reduces a $BDD$ by collapsing identical 
left and right subtrees, and the function {\tt bdd} 
associates this reduced form to $n \in Nat$.
\begin{code}
bdd_reduce (BDD n bt) = (BDD n (reduce bt)) where
  reduce (C b) = C b
  reduce (D _ l r) | l == r = reduce l
  reduce (D v l r) = D v (reduce l) (reduce r)

bdd n = bdd_reduce . plain_bdd n
\end{code}

Note that we omit here the reduction step consisting in
sharing common subtrees, as it is obtained easily by replacing
trees with DAGs. The process is facilitated by the fact
that our unique encoding provides a perfect hashing
key for each subtree. 

The following examples 
show the results returned by {\tt bdd} for {\tt n=2},
with help from printing function {\tt print\_reduced} given
in Appendix.

\begin{codex}
>print_reduced 2
  BDD 2 (C 0)
  BDD 2 (D 1 (D 0 (C 1) (C 0)) (C 0))
  BDD 2 (D 1 (C 0) (D 0 (C 1) (C 0)))
  BDD 2 (D 0 (C 1) (C 0))
  ...
  BDD 2 (D 1 (D 0 (C 0) (C 1)) (C 1))
  BDD 2 (C 1)
\end{codex}

\subsection{From BDDs to Natural Numbers}
One can ``evaluate back'' the binary tree representing the BDD,
by using the pairing function {\tt bitmerge\_pair}.  
The inverse of {\tt plain\_bdd} is implemented as follows:
\begin{code}
plain_inverse_bdd (BDD _ bt) = 
  rshf bitmerge_pair bt

rshf rf (C tt) = tt
rshf rf (D _ l r) = rf ((rshf rf l),(rshf rf r))
\end{code}

\begin{codex}
>plain_bdd 3 42
  BDD 3 
    (D 2 
      (D 1 (D 0 (C 0) (C 0)) 
           (D 0 (C 0) (C 0))) 
      (D 1 (D 0 (C 1) (C 1)) 
           (D 0 (C 1) (C 0))))

plain_inverse_bdd it
  42
\end{codex}
\noindent Note however that {\tt plain\_inverse\_bdd} does not act as an
inverse of {\tt bdd}, given that the {\em structure} of the $BDD$ tree 
is changed by reduction.

\subsection{Boolean Evaluation of BDDs}
This rises the obvious question: how can we recover the original truth
table from a reduced BDD? The obvious answer is: by evaluating it as a
boolean function! The function {\tt ev} describes the $BDD$ evaluator:
\begin{code}
ev (BDD n bt) = eval_with_mask (bigone n) n bt
 
eval_with_mask m _ (C c) = eval_constant m c
eval_with_mask m n (D x l r) = 
  ite_ (var_mn m n x) 
         (eval_with_mask m n l) 
         (eval_with_mask m n r)

eval_constant _ 0 = 0
eval_constant m 1 = m
\end{code}
The function {\tt ite\_} used in {\tt eval\_with\_mask} 
implements the boolean function  {\tt if x then t else e}
using arbitrary length bitvector operations:
\begin{code}
ite_ x t e = ((t `xor` e).&.x) `xor` e
\end{code}
We will use {\tt ite} as the
basic building block for implementing a boolean evaluator for BDDs.

\subsection{The Equivalence}
A surprising result
is that boolean evaluation and structural transformation with
repeated application of
{\em pairing}
produce the same result, i.e. 
the function {\tt ev} also acts as an inverse 
of {\tt bdd} and {\tt plain\_bdd}.

\noindent {\em 
As the following example shows, boolean evaluation {\tt ev}
faithfully emulates {\tt plain\_inverse\_bdd}, 
on both plain and reduced BDDs.
}

\begin{codex}
>plain_bdd 3 42
  BDD 3 
     (D 2 
       (D 1 (D 0 (C 0) (C 0)) 
            (D 0 (C 0) (C 0))) 
       (D 1 (D 0 (C 1) (C 1)) 
            (D 0 (C 1) (C 0))))
ev it
  42
bdd 3 42
   BDD 3 
     (D 2 
       (C 0) 
       (D 1 
         (C 1) 
         (D 0 (C 1) (C 0))))
ev it
  42
\end{codex}

The main result of this subsection can now be summarized as follows:
\begin{prop} \label{tt}
The complete binary tree of depth $n$, obtained by recursive 
applications of {\tt bitmerge\_unpair} on a truth table $tt$
computes an (unreduced) BDD, that, when evaluated, 
returns the truth table, i.e.:
\begin{equation}
plain\_inverse\_bdd~(plain\_bdd~n~tt)) \equiv id
\end{equation}

\begin{equation}
ev~n~(plain\_bdd~n~tt)) \equiv id
\end{equation}

Moreover, {\tt ev} also acts as a left inverse of {\tt bdd}, i.e.

\begin{equation}
ev~n~(bdd~n~tt)) \equiv id
\end{equation}
\end{prop}
\noindent {\em Proof sketch:} The function {\tt plain\_bdd} builds a binary 
tree by splitting the bitstring $tt \in [0..2^n-1]$ up to depth $n$. 
Observe that this corresponds to the Shannon expansion \cite{shannon_all} of the
formula associated to the truth table, using variable order $[n-1,...,0]$.
Observe that the effect of {\tt bitstring\_unpair} is the same as
\begin{itemize}
  \item the effect of {\tt var\_mn m n (n-1)} 
     acting as a mask selecting the left branch, and
\item 
     the effect of its complement, acting as a mask selecting the right
     branch.
\end{itemize}
Given that $2^n$ is the double of $2^{n-1}$, the same invariant holds at each
step, as the bitstring length of the truth table reduces to half. On the other hand,
it is clear that {\tt ev} reverses the action of both {\tt plain\_bdd} and
{\tt bdd}, as BDDs and reduced BDDs represent 
the same boolean function \cite{bryant86graphbased}.

This result can be seen as a yet another intriguing isomorphism between
boolean, arithmetic and symbolic computations.

\section{Ranking and Unranking of BDDs} \label{rank}


One more step is needed to extend the mapping between $BDDs$ with $n$
variables to a bijective mapping from/to $Nat$: 
we will have to ``shift towards infinity'' 
the starting point of each new block of 
BDDs in $Nat$ as BDDs of larger and larger sizes are enumerated.

First, we need to know by how much - so we will count the number
of boolean functions with up to $n$ variables.
\begin{code}
bsum 0 = 0
bsum n | n>0 = bsum1 (n-1)

bsum1 0 = 2
bsum1 n | n>0 = bsum1 (n-1)+ 2^2^n
\end{code}
The stream of all such sums can now be generated as usual\footnote{{\tt bsums}
is sequence A060803 in The On-Line Encyclopedia of Integer
Sequences, \url{http://www.research.att.com/~njas/sequences}}:
\begin{code}
bsums = map bsum [0..]
\end{code}
\begin{codex}
>genericTake 7 bsums
  [0,2,6,22,278,65814,4295033110]
\end{codex}

What we are really interested into, is decomposing {\tt n} into
the distance {\tt n-m} to the
last {\tt bsum} {\tt m} smaller than {\tt n},
and the index that generates the sum, {\tt k}.
\begin{code}
to_bsum n = (k,n-m) where 
  k=pred (head [x|x<-[0..],bsum x>n])
  m=bsum k
\end{code}
{\em Unranking} of an arbitrary BDD is now easy - the index {\tt k}
determines the number of variables and {\tt n-m} determines
the rank. Together they select the right BDD
with {\tt plain\_bdd} and {\tt bdd}.
\begin{code}
nat2plain_bdd n = plain_bdd k n_m
  where (k,n_m)=to_bsum n

nat2bdd n = bdd k n_m
  where (k,n_m)=to_bsum n
\end{code}
{\em Ranking} of a BDD is even easier: we shift its rank
within the set of BDDs with {\tt nv} 
variables, by the value {\tt (bsum nv)} that
counts the ranks previously assigned.
\begin{code}
plain_bdd2nat bdd@(BDD nv _) = 
  (bsum nv)+(plain_inverse_bdd bdd)

bdd2nat bdd@(BDD nv _) = (bsum nv)+(ev bdd)
\end{code}
As the following example shows, {\tt nat2plain\_bdd}
and {\tt plain\_bdd2nat} implement inverse functions.
\begin{codex}
>nat2plain_bdd 42
  BDD 3 
     (D 2 
       (D 1 
         (D 0 (C 0) (C 1)) 
         (D 0 (C 1) (C 0))) 
       (D 1 (D 0 (C 0) (C 0)) 
            (D 0 (C 0) (C 0))))
>plain_bdd2nat it
  42
\end{codex}
\noindent The same applies to {\tt nat2bdd} and its 
inverse {\tt bdd2nat}.
\begin{codex}
>nat2bdd 42
  BDD 3 
     (D 2 
       (D 1 
         (D 0 (C 0) (C 1)) 
         (D 0 (C 1) (C 0))) 
       (C 0))
>bdd2nat it
  42
\end{codex}
\noindent We can now generate infinite streams of BDDs as follows:
\begin{code}
plain_bdds = map nat2plain_bdd [0..]

bdds = map nat2bdd [0..]
\end{code}
\begin{codex}
>genericTake 4 plain_bdds
  [ 
    BDD 0 (C 0),
    BDD 0 (C 1),
    BDD 1 (D 0 (C 0) (C 0)),
    BDD 1 (D 0 (C 1) (C 0))
  ]
genericTake 6 bdds
  [
   BDD 0 (C 0),
   BDD 0 (C 1),
   BDD 1 (C 0),
   BDD 1 (D 0 (C 1) (C 0)),
   BDD 1 (D 0 (C 0) (C 1)),
   BDD 1 (C 1)
  ]
\end{codex}

\section{Multi-Terminal Binary Decision Diagrams (MTBDD)} \label{multi}
MTBDDs \cite{DBLP:journals/fmsd/FujitaMY97,CBGP08} are a natural generalization
of BDDs allowing non-binary values as leaves.
Such values are typically 
bitstrings representing the outputs
of a multi-terminal boolean function,
encoded as unsigned integers.

We shall now describe an encoding of $MTBDDs$
that can be extended to ranking/unranking functions,
in a way similar to $BDDs$ as shown in section \ref{rank}.

Our {\tt MTBDD} data type is a binary tree like the one used for $BDDs$,
parameterized by two integers {\tt m} and {\tt n}, indicating
that an MTBDD represents a function from $[0..n-1]$ to $[0..m-1]$,
or equivalently, an $n$-input/$m$-output boolean function.

\begin{code}   
data MTBDD a = MTBDD a a (BT a) deriving (Show,Eq)
\end{code}

The function  {\tt to\_mtbdd} creates,
from a natural number tt representing a truth table,
an MTBDD representing
functions of type $N \rightarrow M$ with $M=[0..2^m-1], N=[0..2^n-1]$.
Similarly to a BDD, it is represented as binary tree 
of $n$ levels, except that its leaves are in $[0..{2^m}-1]$.
\begin{code}
to_mtbdd m n tt = MTBDD m n r where 
  mlimit=2^m
  nlimit=2^n
  ttlimit=mlimit^nlimit
  r=if tt<ttlimit 
    then (to_mtbdd_ mlimit n tt)
    else error 
      ("bt: last arg "++ (show tt)++
      " should be < " ++ (show ttlimit))
\end{code}
Given that correctness of the range of
{\tt tt} has been checked, the function {\tt to\_mtbdd\_} 
applies {\tt bitmerge\_unpair} 
recursively up to depth $n$, where
leaves in range $[0..mlimit-1]$ are created.
\begin{code}  
to_mtbdd_ mlimit n tt|(n<1)&&(tt<mlimit) = C tt
to_mtbdd_ mlimit n tt = (D k l r) where 
   (x,y)=bitmerge_unpair tt
   k=pred n
   l=to_mtbdd_ mlimit k x
   r=to_mtbdd_ mlimit k y
\end{code}
Converting back from $MTBDDs$ to natural numbers is
basically the same thing as for $BDDs$, except that
assertions about the range of leaf data are enforced.
\begin{code}
from_mtbdd (MTBDD m n b) = from_mtbdd_ (2^m) n b

from_mtbdd_ mlimit n (C tt)|(n<1)&&(tt<mlimit)=tt
from_mtbdd_ mlimit n (D _ l r) = tt where 
   k=pred n
   x=from_mtbdd_ mlimit k l
   y=from_mtbdd_ mlimit k r
   tt=bitmerge_pair (x,y)
\end{code}
The following examples show that {\tt to\_mtbdd} and {\tt from\_mtbdd}
are indeed inverses values in $[0..2^n-1] \times [0..2^m-1]$. 
\begin{codex}
>to_mtbdd 3 3 2008
  MTBDD 3 3 
    (D 2 
      (D 1 
         (D 0 (C 2) (C 1)) 
         (D 0 (C 2) (C 1))) 
      (D 1 
         (D 0 (C 2) (C 0)) 
         (D 0 (C 1) (C 1))))

>from_mtbdd it
2008

>mprint (to_mtbdd 2 2) [0..3]
  MTBDD 2 2 
    (D 1 (D 0 (C 0) (C 0)) (D 0 (C 0) (C 0)))
  MTBDD 2 2 
    (D 1 (D 0 (C 1) (C 0)) (D 0 (C 0) (C 0)))
  MTBDD 2 2 
    (D 1 (D 0 (C 0) (C 0)) (D 0 (C 1) (C 0)))
  MTBDD 2 2 
    (D 1 (D 0 (C 1) (C 0)) (D 0 (C 1) (C 0)))
\end{codex}

\section{Related work} \label{related}
Pairing functions have been used for work on decision problems as early 
as \cite{pepis,kalmar1,robinson50}. 

BDDs are the dominant boolean function representation in
the field of circuit design automation
\cite{DBLP:journals/jcsc/MeinelT99,DBLP:journals/tcad/DrechslerSF04}.

Besides their uses in circuit design automation,
MTBDDs have been used in model-checking and
verification of arithmetic circuits \cite{DBLP:journals/fmsd/FujitaMY97,CBGP08}.

BDDs have also been used in a Genetic Programming context
\cite{DBLP:conf/ices/SakanashiHIK96,DBLP:conf/evoW/2006,DBLP:journals/heuristics/ChenLHW04}
as a representation of evolving individuals subject to crossovers 
and mutations expressed as structural transformations.

\section{Conclusion and Future Work} \label{concl}
Our new pairing/unpairing functions and
their surprising connection to BDDs, 
have been the indirect result of implementation
work on a number of practical applications.
Our initial interest has been triggered by applications of the 
encodings to combinational circuit synthesis \cite{cf08}.
We have found them also interesting as uniform 
blocks for Genetic Programming applications.
In a Genetic Programming context \cite{koza92,poli08}, 
the bijections between bitvectors/natural numbers 
on one side, and trees/graphs representing BDDs on the other side, 
suggest exploring the mapping and its action on various
transformations as a phenotype-genotype connection. 
Given the connection between BDDs to
boolean and finite domain constraint solvers
it would be interesting to explore in that context,
efficient succinct data representations
derived from our BDD encodings.

\bibliographystyle{plainnat}
\bibliography{INCLUDES/theory,tarau,INCLUDES/proglang,INCLUDES/biblio,INCLUDES/syn}

\subsection*{Appendix}
To make the code in the paper fully self contained, 
we list here some auxiliary functions.\\

\noindent {\bf Bitvector Boolean Operation Definitions}
\begin{code}
type Nat=Integer

nand_ :: Nat->Nat->Nat->Nat
nor_ :: Nat->Nat->Nat->Nat
impl_ :: Nat->Nat->Nat->Nat
less_ :: Nat->Nat->Nat->Nat

nand_ mask x y = mask .&. (complement (x .&. y))
nor_ mask x y = mask .&. (complement (x .|. y))
impl_ mask x y = (mask .&. (complement x)) .|. y
less_ _ x y = x .&. (complement y)
\end{code}

\noindent {\bf Boolean Operation Encodings and Names}
\begin{code}
-- operation codes
opcode m 0 = nand_ m
opcode m 1 = nor_ m
opcode m 2 = impl_ m
opcode m 3 = less_ m
opcode _ 4 = xor
opcode _ n = error ("unexpected opcode:"++(show n))

-- operation names
opname 0 = "nand"
opname 1 = "nor"
opname 2 = "impl"
opname 3 = "less"
opname 4 = "xor"
opname n = error ("no such opcode:"++(show n))
\end{code}

\noindent {\bf A Few Interesting Libraries}
\begin{code}
mixops = [0,2]
symops = [0,1]
asymops = [2,3]
\end{code}

\noindent {\bf Tests for the Circuit Synthesizer}
\begin{code}
t0 = findFirstGood symops 3 8 71
t1 = syn asymops 3 71
t2 = mapM_ print (synall mixops 2)
t3 = syn asymops 3 83 -- ite
t4 = syn symops 3 83
t5 = syn [0..4] 3 83 -- ite with all ops
-- x xor y xor z -- cpu intensive
t6 = syn asymops 3 105 
\end{code}

\noindent {\bf Bit crunching functions} 

\noindent This function splits a natural number in a set of natural
numbers indicating the positions of its {\tt 1} bits in its right to left binary
representation.
\begin{code}
nat2set n = nat2exps n 0 where
  nat2exps 0 _ = []
  nat2exps n x = 
    if (even n) then xs else (x:xs) where
      xs=nat2exps (div n 2) (succ x)
\end{code}
\noindent This function aggregates a set of natural
numbers indicating positions of {\tt 1} bits 
into the corresponding natural number.
\begin{code}
set2nat ns = sum (map (2^) ns)
\end{code}

\noindent {\bf I/O functions}

\noindent These functions print out the BDDs
of all the $2^{2^k}$ truth tables 
associated to $k$ variables.
\begin{code}
print_plain k = mapM_ 
  (print . (plain_bdd k)) [0..(bigone k)] 
print_reduced k = mapM_ 
  (print . (bdd k)) [0..(bigone k)] 
\end{code}
\noindent This function applies f to a list of objects and prints the results
on successive lines.
\begin{code}   
mprint f = (mapM_ print) . (map f)
\end{code}

\end{document}